\documentclass{jfm}

\usepackage{graphicx}
\usepackage{natbib}
\usepackage{upmath}
\usepackage{amssymb}
\usepackage{amsbsy}

\title[Aeroelastic instability of cantilevered flexible plates in uniform flow]
{Aeroelastic instability of cantilevered \break flexible plates in uniform flow}

\author[C. Eloy, R. Lagrange, C. Souilliez and L. Schouveiler]%
{C\ls H\ls R\ls I\ls S\ls T\ls O\ls P\ls H\ls E\ns  E\ls L\ls O\ls Y\ls ,\ns
R\ls O\ls M\ls A\ls I\ls N\ns  L\ls A\ls G\ls R\ls A\ls N\ls G\ls E\ls ,\break
C\ls L\ls A\ls I\ls R\ls E\ns  S\ls O\ls U\ls I\ls L\ls L\ls I\ls E\ls Z\ns
\and
L\ls I\ls O\ls N\ls E\ls L\ns  S\ls C\ls H\ls O\ls U\ls V\ls E\ls I\ls L\ls E\ls R}

\affiliation{IRPHE, CNRS \& Aix-Marseille Universit\'e,
49 rue Joliot-Curie, 13013 Marseille, France}

\begin{document}

\maketitle

\begin{abstract}
We address the flutter instability of a flexible plate immersed in an axial flow. 
This instability is similar to flag flutter and results from the competition 
between destabilising pressure forces and stabilising bending stiffness. 
In previous experimental studies, the plates have always appeared
much more stable than the predictions of two-dimensional models.
This discrepancy is discussed and clarified in this paper
by examining experimentally and theoretically
the effect of the plate aspect ratio on the instability threshold. 
We show that the two-dimensional limit cannot be achieved experimentally
because hysteretical behaviour and three-dimensional effects appear for plates of large aspect ratio.
The nature of the instability bifurcation (sub- or supercritical) is also discussed 
in the light of recent numerical results.  
\end{abstract}

\section{Introduction}

The flutter of a flexible plate immersed in an axial flow is a
canonical example of flow-induced vibrations. This instability can
be experienced in everyday life when one observes a flag flapping in
the wind. Because this phenomenon appears in many applications (paper
industry, airfoil flutter, snoring), it has motivated a
large literature which has been recently reviewed by \cite{Paidoussis_book2}. 
This instability can be regarded as a competition between fluid forces and
elasticity. Indeed, when the plate experiences a small lateral
deflection, a destabilising pressure jump can appear across the plate, 
while the bending stiffness tends to bring the plate back to the stable planar state.

This system can be studied by restricting the
analysis to one-dimensional flutter modes as observed in most experiments. 
In this case, the plate motion obeys
the Euler-Bernoulli beam equation with additional pressure forces 
which are calculated by assuming a potential flow. 
To simplify the problem further,
\cite{Shelley2005} considered a plate infinite 
in both directions in a similar way to the stability
analysis of a jet by Lord \cite{Rayleigh79} who already noted
in his seminal paper the analogy with the problem of
flag flutter.

In other theoretical studies, the plate length
$L$ (or chord) takes a finite value while two asymptotic limits are
considered for its span $H$. If $H\ll L$, the fluid forces can be
calculated using the slender body theory of \cite{Lighthill60} \citep[e.g.][]{Datta75,Lemaitre2005}
 and if $H\gg L$ the problem can be treated
as two-dimensional \citep[as done by][]{Kornecki76,Huang95,Watanabe2002b,Guo2000}. 
In this latter case, the flow
is entirely described by point-vortices which are distributed within
the plate and possibly in its wake. It is known from airfoil
theory that this problem does not admit a unique solution
(intrinsically because the Laplace equation has to be solved on                                   
an open domain). 
\cite{Kornecki76} used two approaches to treat this non-unicity.
They first considered that the total circulation
around the plate vanishes and thus no vorticity is shed in the wake. 
Second, they used the Kutta condition applied at the
trailing edge to prescribe the circulation around the plate 
as \cite{Theodorsen1935} did in his study of airfoil flutter 
\citep[see also][ for an application of the Kutta hypothesis to unsteady flows]{Crighton1985}. 
The Kutta condition imposes
advected vortices in the wake which makes this model more physical
for this flag-type instability 
\cite[see for instance chapter 5-6 of][]{Bisplinghoff1983}. 
Note that this shed vorticity can
equivalently be regarded as the consequence of the Kelvin's
circulation theorem.
Using these two-dimensional flow models, a stability analysis can be
carried out that permits the prediction of the critical velocity for plate
flutter. 

The first visualisations of the flutter instability have been performed 
with flags made of various fabrics by \cite{Taneda68}. 
A few years later, \cite{Datta75} carried out experiments with long 
ribbons hanging in airflow. This slender body limit ($H\ll L$) has been
reexamined recently by \cite{Lemaitre2005} and results show good
agreement with linear stability analysis. 
For larger aspect ratios, \cite{Huang95}, \cite{Yamaguchi2000} and 
\cite{Watanabe2002a} have considered both
the effects of the plate length and of the material properties 
on the flutter instability. However, to the our knowledge, 
the effect of aspect ratio has not been investigated up to now. 

Cantilevered plates in axial flow have also been modelled 
numerically. \cite{Watanabe2002b} and \cite{Balint2005} 
used a two-dimensional Navier-Stokes solver combined to 
a linear beam model for the plate.
The critical velocities found in these studies agree with the
results of two-dimensional stability analysis.
The nature of the bifurcation in this flutter instability has also been studied
using a nonlinear beam equation for the plate and 
two-dimensional \citep{Tang2007} or three-dimensional \citep{Tang2003}
vortex methods to model the potential flow. 
Similarly to the slender body limit \citep{Yadykin2001}, the bifurcation is shown to be
supercritical in these models. However, \cite{Alben2008} recently found hysteresis and bistability
using a  two-dimensional model taking into account the nonlinearities originating both from 
the flow and from the elastic plate.
Hysteresis has
also been reported in previous experiments with an hysteresis loop at least an order of magnitude
wider than in the simulations of \cite{Alben2008}. This indicates a subcritical bifurcation.
To date, the real nature of the bifurcation and the apparent discrepancy between most numerical
models and the experiments remains unexplained.

The results of existing two-dimensional models have been compiled by
\cite{Watanabe2002b} and compared with existing experimental data
\citep[see also the recent comparison made by][]{Tang2007}. 
Remarkably  all these theories predict
approximately the same critical velocities but strongly
underestimate the measured thresholds. In other words, the plate
appears systematically more stable than predicted by a
two-dimensional approximation. This discrepancy has motivated the
recent study of \cite{Eloy2007} in which the
finite plate span is explicitly taken into account in the analysis. 
\cite{Shayo80} already
addressed the effect of aspect ratio on the flutter instability. However, he made
several mathematical assumptions to simplify the stability analysis which led
him to falsely conclude that the larger the aspect ratio is the more stable is the system.
The present
paper aims at comparing the predictions of \cite{Eloy2007} with experimental
measurements in which the effect of the plate aspect ratio is
extensively investigated. 

This paper is divided as follows: in~\S\,\ref{sec:2} the physical model is presented
and the dimensionless parameters are introduced; in~\S\,\ref{sec:3} the experimental
setup is briefly described and the main results are given in~\S\,\ref{sec:4}; 
finally these results are discussed in the light of previous theoretical and numerical
models in~\S\,\ref{sec:5}.

\section{Physical model}
\label{sec:2}

\begin{figure}
\centerline{
\includegraphics[height=3.3cm]{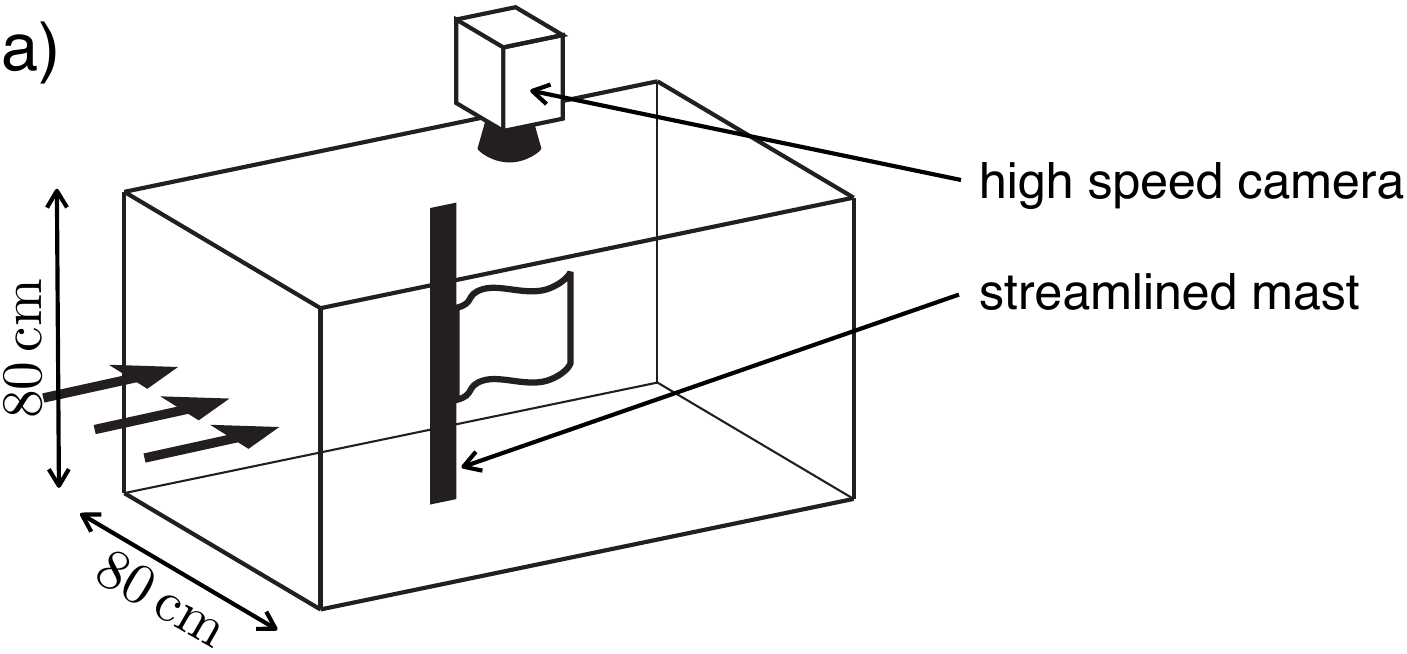} ~~~
\includegraphics[height=3.3cm]{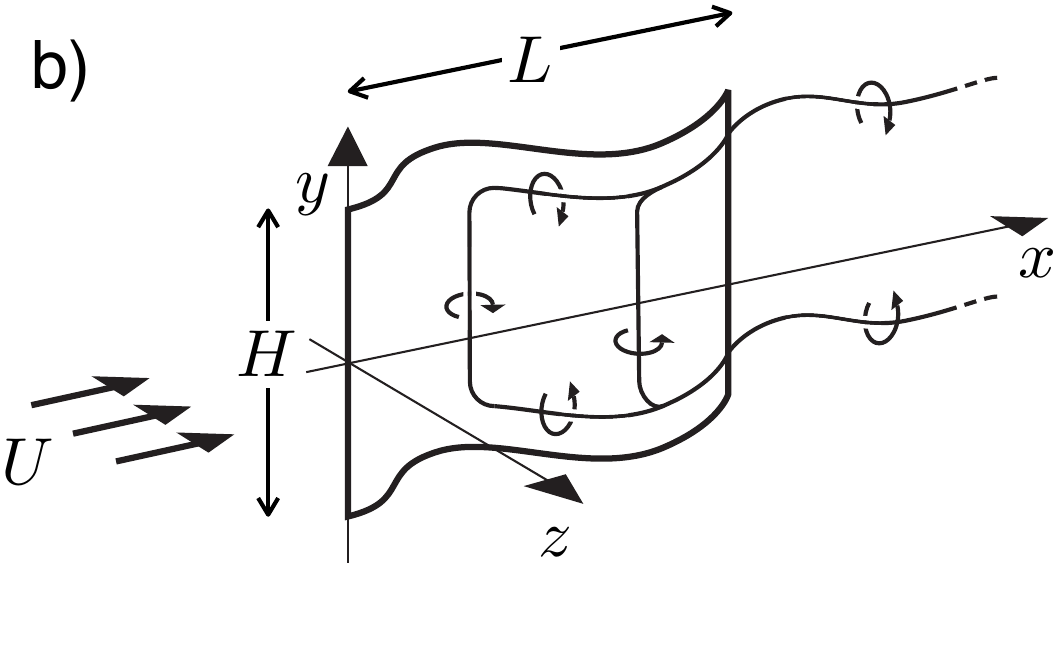}
}
\caption{\label{fig:sketch} Sketches of the experimental setup (a) and of the flexible 
plate subject to the one-dimensional deflection $z(x,t)$ (b).}
\end{figure}

As shown in figure~\ref{fig:sketch}, we consider a flexible plate of span $H$ and length $L$, 
lying in the vertical plane $(Oxy)$ and immersed in an axial flow of
velocity $U$. Its flexural rigidity is given by $D=E h^3/12(1-\nu^2)$, where $E$ is the 
Young's modulus, $h$ the plate thickness and $\nu$ its Poisson's ratio.
For small lateral deflections $z(x,t)$, 
the plate motion is driven by the linearised Euler-Bernoulli equation
\begin{equation}
m \,\partial_{t}^2 z +D\,\partial_{x}^4 z+\langle p \rangle =0,
\label{eq:Euler}
\end{equation}
where $m$ is the mass per unit surface of the plate, $p(x,y,t)$ is the pressure jump
across the plate and the notation $\langle . \rangle$ stands for the average along the span $H$. 
Equation~(\ref{eq:Euler}) is valid in the limit of an inviscid fluid and for a material of
negligible visco-elastic damping. In addition the deflection $z$ satisfies clamped boundary conditions at the
leading edge: $z(0)=\partial_{x}z(0)=0$ and free boundary conditions at the trailing edge:
$\partial^2_{x}z(L)=\partial^3_{x}z(L)=0$.

Assuming an inviscid flow, the perturbation velocity can be fully described 
by the vorticity distribution in the flow. Vortex-lines are located in the plate and in its wake
as sketched in figure~\ref{fig:sketch}(b).
The pressure jump across the plate only depends on the $y$-component
of the vorticity distribution $\gamma(x,y,t)$ through the unsteady Bernoulli equation
\begin{equation}
\partial_{x} p=\rho (\partial_{t} +U\partial_{x})\gamma,
\label{eq:Bernoulli}
\end{equation}
where $\rho$ is the fluid density and $\gamma$ has the dimension of a velocity 
(i.e. a circulation per unit length). 
As seen from this equation a vorticity distribution of the form 
$\gamma_{0}(y)\exp[\mathrm{i}(k x- \omega t)]$
with wave velocity $\omega/k=U$ is compatible with a zero pressure
jump and is indeed the distribution of vorticity in the wake for small
plate deflections.

In order to couple the vorticity distribution $\gamma$ to the plate deflection $z$,
a kinematic condition is enforced on the plate surface which can be written as
\begin{equation}
C \hspace{-11pt}\int \langle \gamma(\xi) \rangle K(x-\xi) \mathrm{d} \xi=  
	(\partial_{t} +U\partial_{x})z = w(x,t),
	\label{eq:inversepb}
\end{equation}
where the function $K$ is a kernel defined below,
$\xi$ is a dummy variable in place of $x$, $w$ is the normal plate velocity in the fluid reference  
frame,  the brackets $\langle . \rangle$ still denote averaging along $H$ and  the letter 
$C$ across the integral sign indicates that the Cauchy principal value should be taken. 
Since this inverse problem can have non-unique solutions, the Kutta condition is also
used at the trailing edge i.e. $p(L^{-})=0$. 
The kernel $K$ in (\ref {eq:inversepb})  expresses the influence of a vertical vorticity line located 
in $\xi$ on the $z$-component of the fluid velocity in $x$. For an infinite span, the azimuthal velocity of
a point-vortex in two dimensions yields the kernel $K(X)=1/2\pi X$
used in all the two-dimensional models. 
For an asymptotically small $H$ the Lighthill's slender body theory
is equivalent to taking $K(X)=\mbox{sgn}(X)/\pi H$. In \cite{Eloy2007}, we have shown
that an approximate kernel can be used for any $H$
\begin{equation}
K(X)=\frac{1}{2 \pi X} +\left( \frac{1}{\pi H} - \frac{1/\pi-1/8}{H+|X|}\right) \mbox{sgn}(X).
\label{eq:K}
\end{equation}
This kernel is constructed as a composite of the correct asymptotic expansions for small and large $H$. 
Morevover, for all $X$, it 
has been shown to be within $2\%$ of the exact kernel which cannot be expressed analytically.

Applying the operator $\rho (\partial_{t} +U\partial_{x})$ on (\ref{eq:inversepb}) and integrating by parts yields the following inverse problem for the pressure jump
\begin{equation}
C \hspace{-11pt}\int \langle \partial_{\xi}p(\xi) \rangle K(x-\xi) \mathrm{d} \xi=  
	\rho(\partial_{t} +U\partial_{x})^2 z=\rho(\partial_{t} +U\partial_{x}) w.
	\label{eq:inversepbP}
\end{equation}
This Fredholm equation of the first kind has been solved in \cite{Eloy2007} 
to address the flutter instability of the plate. 
But, since an integration by parts was
performed to obtain (\ref {eq:inversepbP}), 
the pressure jump is implicitly assumed to be non-singular at
leading and trailing edges. 
At the trailing-edge, this assumption is correct because the singularity 
vanishes thanks to the Kutta condition.
However, one expects an inverse square-root pressure singularity at the leading-edge.  
As we shall see in \S4, the instability characteristics are not greatly modified by 
the treatment of this leading edge singularity. It is probably due to the fact that  
it corresponds to the clamped end of the plate where pressure forces do no work. 
The model derived from \cite{Eloy2007} will be called \emph{model without singularity} 
in the rest of the paper.

To assess the validity of this model  and to allow comparisons with the two-dimensional model of \cite{Kornecki76}, another approach has been used in the present paper. 
It is valid in the limit of large aspect ratio and it will be referred to as the
\emph{asymptotic model}. 
In this model, the two leading order terms have been retained in (\ref{eq:K}) leading to
the following kernel
\begin{equation}
K(X)=\frac{1}{2 \pi X} +\frac{1}{8H}\mbox{sgn}(X) + O (H^{-2}).
\label{eq:K2}
\end{equation}
Note that we have shown in \cite{Eloy2007} that this  is the mathematically 
correct expansion for $H \gg |X|$. 
The problem being linear by construction, the stability of this  fluid-structure coupling
can be now addressed assuming a Galerkin decomposition for the deflection
\begin{equation}
z(x,t)=\sum^{\infty}_{n=1} a_{n}z_{n}(x)\mathrm{e}^{\mathrm{i}\omega t},
\end{equation}
where the $z_{n}$ are the orthogonal eigenmodes of a beam \textit{in vacuo} thus satisfying
the proper boundary conditions.  
For a given Galerkin mode, one has then to invert (\ref{eq:inversepb}) with 
$K$ given by (\ref{eq:K2}) 
and insert the solution in (\ref{eq:Bernoulli}) to find the corresponding pressure
jump $\langle p_{n} \rangle \exp (\mathrm{i}\omega t)$. 
This procedure allows the leading edge singularity to be taken into account properly.
The partial differential equation (\ref{eq:Euler}) is then reduced to an eigenvalue problem
for the complex frequency $\omega$ whose eigenmodes 
correspond to the instability modes. 
For small flow velocities, all modes are stable (i.e. their complex frequencies $\omega$ 
have a positive imaginary part). When $U$ is larger than a critical flow velocity $U_{c}$, 
one of the instability modes eventually becomes unstable. 

Using $L$ and $L/U$ as characteristic length and time, the system parameters are reduced to
three dimensionless numbers: 
the reduced velocity $U^*$, the mass ratio $M^*$ and the aspect ratio $H^*$ given by
\begin{equation}
U^*= LU\sqrt{m/D}, \quad M^*=\rho L /m, \quad H^*=H/L.
\end{equation}
With these dimensionless parameters, 
the present `asymptotic model' and the `model without singularity' detailed in \cite{Eloy2007} 
permit the prediction of the critical velocity 
$U^*_{c}(M^*,H^*)$ and the dimensionless mode frequencies $\omega^*=\omega L/U$. 
The results of the two-dimensional model of 
\cite{Kornecki76} have also been 
recalculated in the present paper with better computer accuracy to allow comparisons. It is
strictly equivalent to the `asymptotic model' for infinite span $H$.


\section{Experimental setup}
\label{sec:3}

As sketched in figure~\ref{fig:sketch}(a) experiments were performed in a
low-turbulence closed wind tunnel of $80\times 80\;$cm$^2$ cross section. 
The wind velocity $U$ could be
varied continuously up to $65\;$m$\,$s$^{-1}$ and was measured with a Pitot tube.
The plates were clamped in a streamlined mast crossing the wind tunnel vertically.
To ensure that the mast had negligible effects on the instability,  
two masts of different cross section were used 
(mast 1: thickness$\times$chord$=4\;$mm$\times 20\;$mm; mast 2: $9\;$mm$\times 43\;$mm).
Plates were cut in Mylar sheets of
mass per unit area $m=0.14\;$kg$\,$m$^{-2}$ and flexural rigidity
$D=0.48\times 10^{-3}\; $N$\,$m. To estimate $D$,
the deflection due to gravity was measured for horizontally clamped Mylar strips of
various lengths. The plate length $L$ was varied between
$2$ and $30\,$cm for a thickness $h=0.16\,$mm.

In all experiments presented in this paper, 
the same protocol has been followed. The
plate is clamped in the mast and the flow velocity is slowly
increased starting from zero. At small velocities the plate appears stable, i.e. 
steady and aligned with the flow. Eventually, for a critical flow
velocity $U^*_c$, the plate flutters spontaneously with a large amplitude
and a well-defined frequency.
Then the flow velocity is slowly decreased in small decrements until the
plate returns to its stable state again at another critical
velocity $U^*_d$ as illustrated in figure~\ref{fig:hysteresis}. 

Visualisations were carried out through the top wall of the wind tunnel with a
high speed video camera aligned with the $y$-axis (see
figure~\ref{fig:sketch}a). 
The top edge of the plate was painted white and illuminated
to record its motion and aid measuring the flutter frequency.  
In the present experiments, the camera was operating at 
$300\,$Hz with a $512\times 512$ pixel resolution 
and the exposure time was varied between $300$ and $1500\;\mu$s.
The camera visualisations are used to extract the flutter 
amplitude $A$  and its angular frequency $\omega^*$. 
This is done by detecting in each snapshot of the plate deflection at $3/4$ of the total 
plate length.


\section{Results}
\label{sec:4}

\begin{figure}
\centerline{\includegraphics[width=0.6\textwidth]{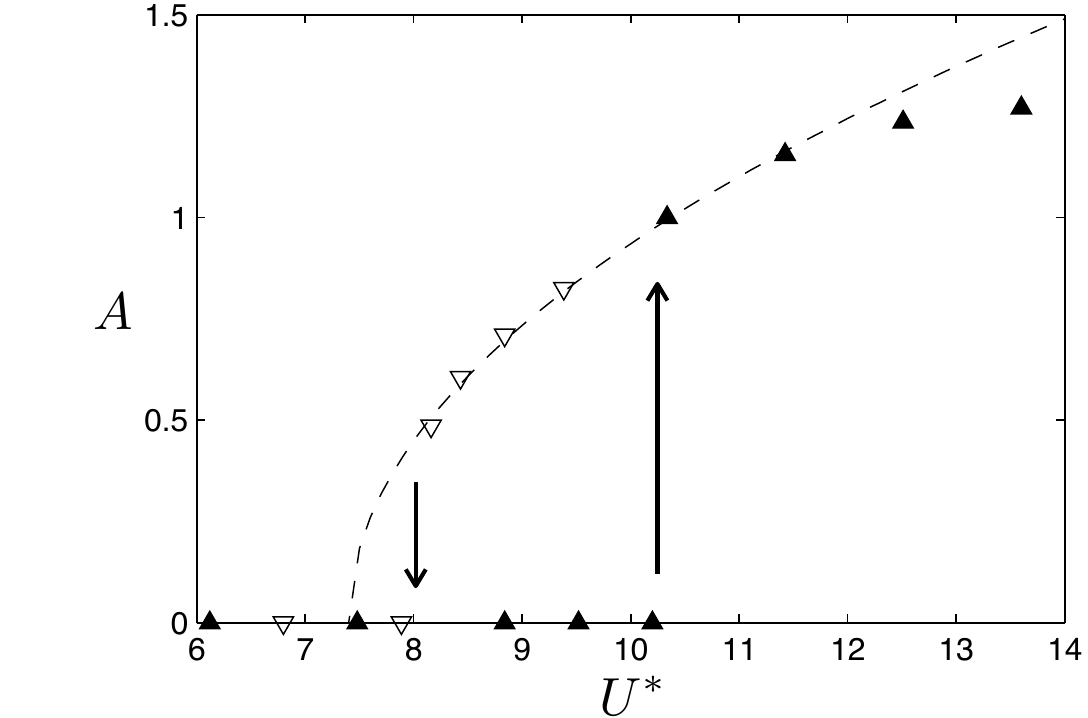}}
\caption{\label{fig:hysteresis} Mode amplitude $A$ (in arbitrary units) as a function of the
reduced velocity $U^*$. Amplitude is plotted as filled triangles when velocity is increased and 
open triangles when it is decreased.
The experimental parameters are $H^*=1$ and $M^*=0.6$. The
instability exhibits a strong hysteresis with $U^*_c=10.3$ and $U^*_{d}=8$. 
However, the amplitudes are well fitted by the square-root law 
 $A=58(U^*-7.4)^{1/2}$ (dashed line) suggesting a supercritical bifurcation.}
\end{figure}
 
The flutter amplitude $A$
is plotted in figure~\ref{fig:hysteresis} as a function of the reduced velocity; this plot exhibits a
hysteretic  behaviour. This has been observed in all experiments with
sufficiently large aspect ratio ($H^* \gtrsim 1$). This hysteresis together with the large mode amplitude
observed at threshold evokes a subcritical bifurcation as recently
suggested by the numerical results of \cite{Alben2008}. 
Note however that the wideness of the hysteresis loop observed in the present experiments is of the order
of 20\% (it is defined as $(U^*_{c}-U^*_{d})/U^*_{c}$) 
whereas it is only of the order of 1\% in the simulations of \cite{Alben2008}.
In other numerical models \citep{Yadykin2001,Tang2003,Tang2007}, the bifurcation is found to be supercritical. 
Figure~\ref{fig:hysteresis}
shows that the amplitudes near threshold can be reasonably well fitted with a square-root law giving 
evidence that the bifurcation could indeed be supercritical and that the observed hysteresis could just be
an artifact as discussed below.

\begin{figure}
\centerline{\includegraphics[width=0.6\textwidth]{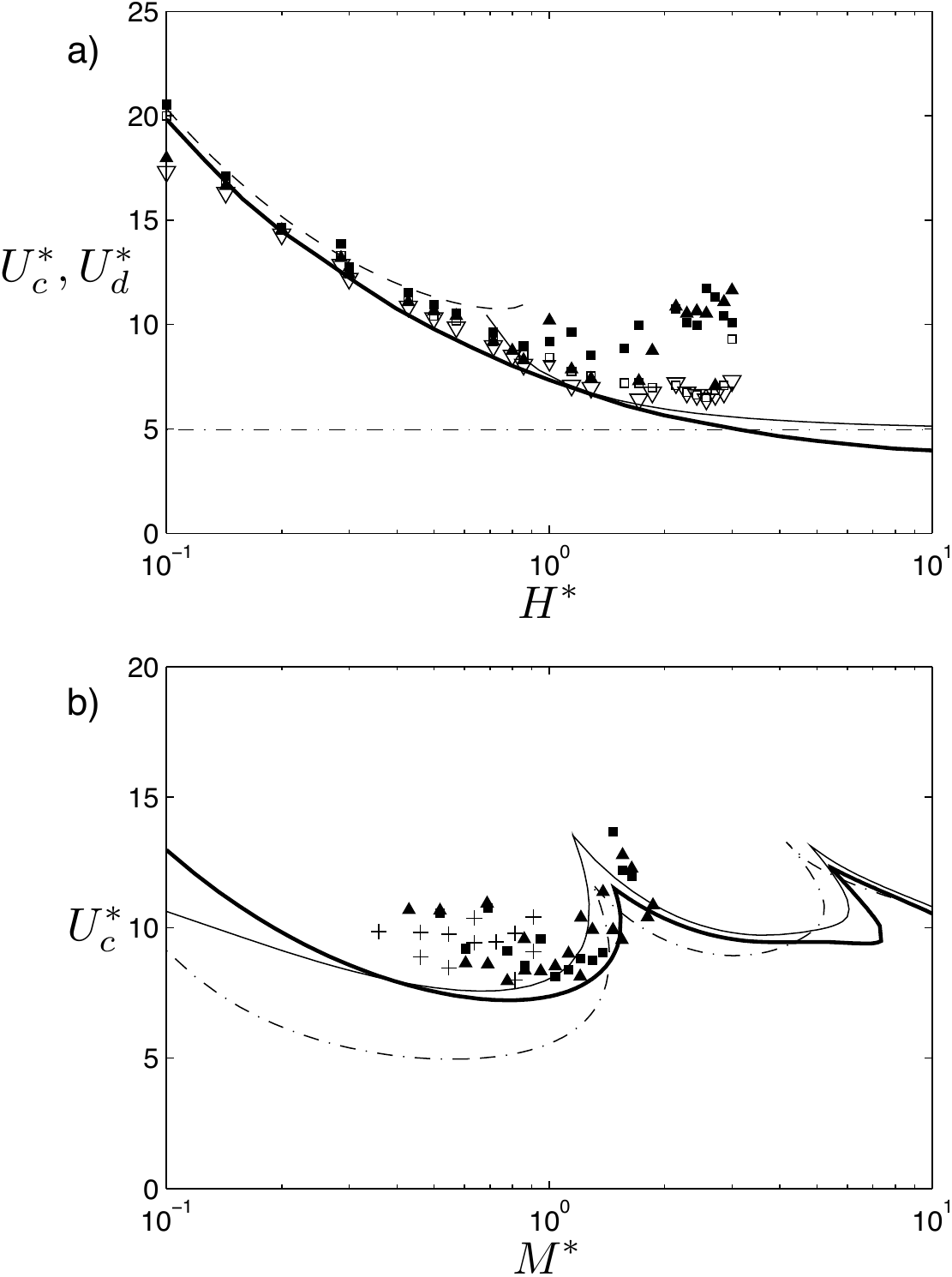}}
\caption{\label{fig:criticalU} Critical velocities $U^*_c$ and $U^*_d$ as a function of the aspect ratio 
$H^*$ for $M^*=0.6$ (a) and as a function of the mass ratio $M^*$ for
$H^*=1$ (b). Filled squares and triangles correspond to measured $U^*_c$ for the masts 1 and 2 
respectively and the same open symbols refer to $U^*_{d}$. Crosses correspond to the experiments
of \cite{Huang95} for $0.6<H^*<1.5$. The thick solid line is the three-dimensional `model
without singularity' \citep{Eloy2007} and the thin solid line is the `asymptotic model' of 
the present paper for the same parameters. The dashed line is the slender body theory
\citep[similar to][]{Lemaitre2005} and the dash-dotted line corresponds to the two-dimensional theory 
with circulation \citep[same as][with better computer accuracy]{Kornecki76}. } 
\end{figure}

The critical velocities $U^*_c$ and $U^*_d$ are
plotted in figure~\ref{fig:criticalU} as a function of the aspect ratio $H^*$ and the mass ratio $M^*$.
First the plate span is varied with the plate length remaining constant. This allows the aspect
ratio to be varied without changing the other parameters. As expected from slender body theory,
the critical velocity tends to decrease as aspect ratio is increased. 
From figure~\ref{fig:criticalU}(a) two regimes can be distinguished. For small aspect ratio ($H^* \lesssim 1$),
the measured critical velocities are well predicted by the linear stability analysis and the hysteresis is
very small if not absent. 
For larger aspect ratio ($H^* \gtrsim 1$) the hysteresis greatly increases. The experimental 
data points are also more scattered.  For these large aspect ratios,
the linear stability analysis tends to underestimate the critical 
velocity $U^*_{c}$. Moreover, this threshold seems to slightly increase for $1<H^*<3$ contrarily to the
decreasing threshold $U^*_d$ which remains close to the predictions. 

The figure~\ref{fig:criticalU}(a) also permits to quantify the error made by the `model without
singularity'. Indeed it can be seen that it does not exactly converge to the two-dimensional limit 
when aspect ratio tends to infinity but slightly below. As explained above, this is because in this model 
the pressure jump is assumed to be regular at the leading edge whereas, in the `asymptotic model'
and in the two-dimensional model of \cite{Kornecki76}, the pressure jump has the physically correct 
inverse square root singularity.

\begin{figure}
\centerline{\includegraphics[width=0.65\textwidth]{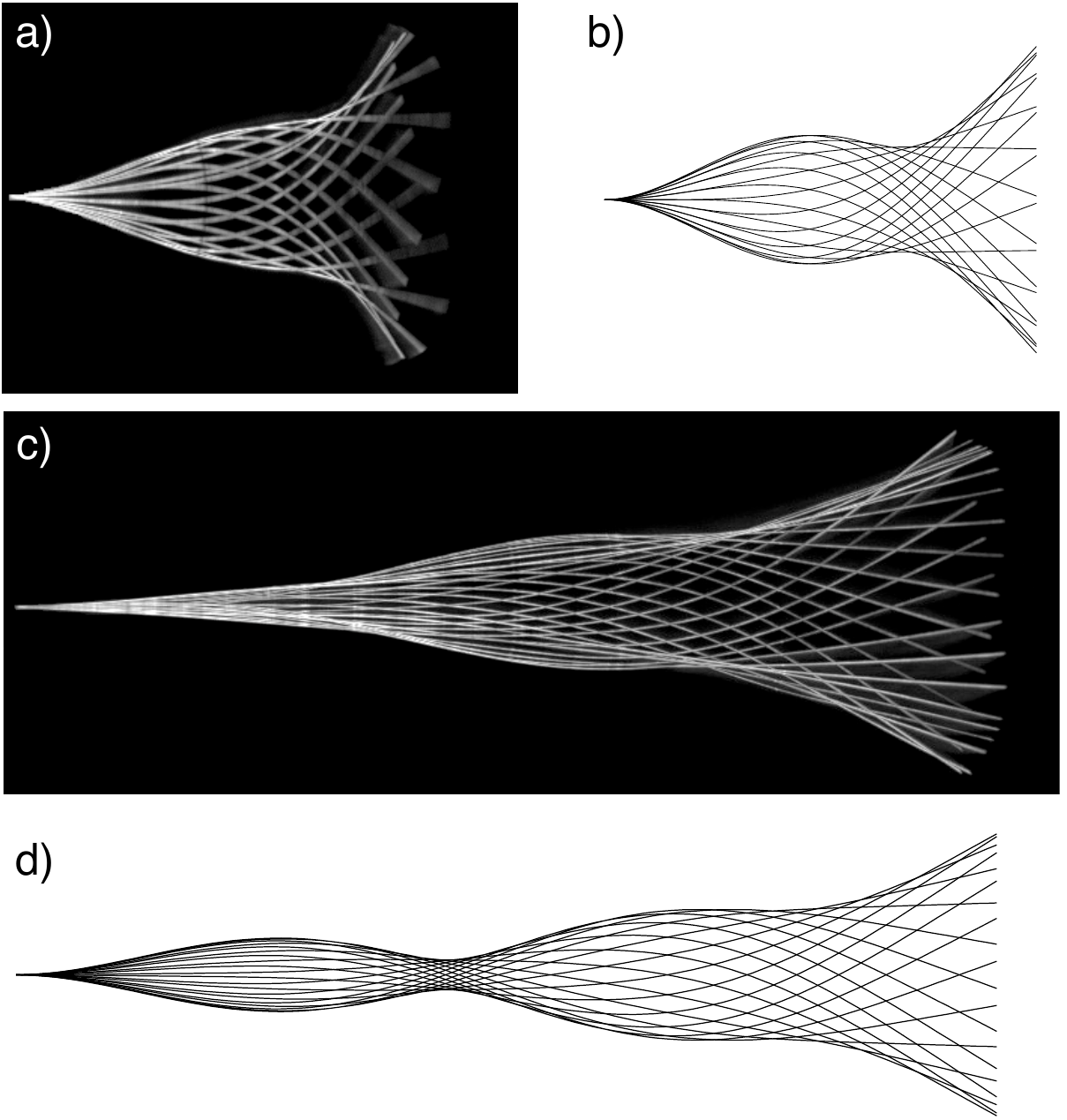}}
\caption{\label{fig:modes} Superimposed views of the flutter modes for two plates of same
aspect ratio $H^*=1$ but with different sizes [$M^*=0.74$ for (a) and (b), $M^*=1.94$
for (c) and (d)]. These amplitudes are obtained at threshold and the measured frequencies are
$\omega^*=1.82$ for $U^*_{c}=8.1$ (a) and $\omega^*=2.5$ for $U^*_{c}=10.9$ (c). The 
experimental snapshots are compared with the modes predicted by the present linear stability analysis (with
arbitrary amplitude): $\omega^*=2.4$ for $U^*_{c}=7.2$ (b) and $\omega^*=3.7$ for $U^*_{c}=10.4$ (d).}
\end{figure}

In figure~\ref{fig:criticalU}(b), the aspect ratio has been kept constant ($H^*=1$) by varying by the same
factor the plate length and span. 
The measured critical velocity $U^*_{c}$ is compared to the present theoretical predictions
and to the two-dimensional model of \cite{Kornecki76}. 
Experimental data from \cite{Huang95} are also reported even if these experiments have been
carried out with plates spanning on the entire tunnel width with a $2\,$mm clearance. This experimental
setup was expected to  model a two-dimensional flow.
In the present analysis, the finiteness of the span is taken into account 
leading to a smaller pressure jump $\langle p \rangle$ in (\ref{eq:Euler}) 
compared to the two-dimensional model and thus a better prediction of the instability threshold.
The three models permit to predict different instability modes as $M^*$ is increased as illustrated by the different
lobes in figure~\ref{fig:criticalU}(b). 
If modes are numbered by order of ascending frequencies, the mode two is observed 
for the smallest mass ratios
($M^* \lesssim 1.5$). This single-neck mode is pictured in 
figure~\ref{fig:modes}(a,b). For larger mass ratios  ($1.5 \lesssim M^* \lesssim 5$), the mode three, a double-neck mode,
is the first unstable one as illustrated in figure~\ref{fig:modes}(c,d), 
and for even larger mass ratios ($M^* \gtrsim 5$), one expect to observe higher order modes. 
Note that the mode one (the mode with the smallest frequency) 
is never unstable for a clamped-free plate 
and accordingly never observed in the experiments
\citep[as discussed in][]{Guo2000}.
In other words, as $M^*$ increases, the typical mode wavelength $\lambda$ is a smaller fraction of the plate
length $L$ and therefore if the aspect ratio is kept constant, $H/\lambda$ increases. This span to
wavelength ratio has to be large for a two-dimensional theory to be valid and
this explain why the difference between the results of the two- and three-dimensional theories 
reduces as $M^*$ increases in figure~\ref{fig:criticalU}(b).

In figure~\ref{fig:modes}(a,c), superimposed views of the
plate recorded with the camera during one flutter period are shown.  
These visualisations were carried out at the instability threshold $U^*_{c}$ for plates
of same aspect ratio $H^*=1$ but of two different mass ratios. 
As predicted by the analysis, two different modes are observed as $M^*$ varies.
From these data, the mode shapes and their frequencies can be compared to the theoretical
predictions for the same parameters.
However nonlinear effects are probably important
given the large amplitude observed at threshold and 
the agreement with the results of the linear stability 
analysis are only qualitative.


\section{Discussion}
\label{sec:5}

In this paper, we have studied experimentally and theoretically the flutter instability
of cantilevered flexible plates in uniform flow. We have shown in particular that the
three-dimensionality of the flow has to be taken into account to predict accurately
the instability threshold as the plate aspect ratio is varied. We have also shown that
hysteresis is present for plates of large aspect ratio and that the instability
threshold measured when increasing the velocity does not converge towards the
two-dimensional limit as aspect ratio tends to infinity. 
These experimental results are now discussed with regard to the assumptions 
made in the stability analysis.

In the experiments with large aspect ratio ($H^*\gtrsim 2$), we have observed that
the flutter mode is no longer purely one-dimensional as the top corner
flutters with a larger amplitude than the rest of the plate. 
Moreover, in some experiments with a very large aspect ratio ($H^*=4$) and
for flow velocity just below threshold, 
the same top corner flutters while the rest of the plate is still
motionless. These two-dimensional motions indicate 
that the Euler-Bernoulli beam equation is no longer valid 
for the plate and one should use the nonlinear
F\"oppl--von K\'arm\'an equations as soon as the deflection along the span
is of the order of the plate thickness \cite[see for instance][]{Landau1986}. 
Several reasons may account for the two-dimensional deflections of the plate. 
First, the pressure field over the 
plate is non uniform along the span and exhibit a maximum at
mid-span and zeros at the plate edges. Second, the gravity field
induces a non trivial stress tensor in the plate that results in compression
at the top corner and tension at the bottom corner. Since tension has
a stabilising effect, this may explain why the plate deflection exhibits a larger
amplitude at the top corner for large aspect ratios. 

The two-dimensional plate deflections can also originate from 
small imperfections in the flow 
caused by the wind tunnel or the mast wake. 
These imperfections would create  flow unsteadiness 
and two-dimensional plate vibrations could 
be forced. A small
deflection along the span could also be caused by imperfections of the plate clamping or planarity. 
But whatever their origin,
as soon as these two-dimensional deflections or vibrations along the span are of the
order of the plate thickness, they act as a stiffening effect.  
Gauss curvature being energetically costly, once plate is bent along its span,
it will bend more difficultly along its length and the instability will be delayed. 
However, once the plate is unstable, 
its flutter mode can be purely one-dimensional, flattening any bending along the span, thus
generating a hysteresis loop. 
This could be an elegant explanation 
for the appearance of the hysteresis loop only for aspect ratios larger than one. Indeed, in this
case, the elastic energy required to bend the plate along its span becomes smaller 
than the energy to bend it along its length. It is then reasonable to think that any 
defect may trigger deflections along the span of the order of the plate thickness and thus hysteresis. 

If one assumes that hysteresis is due to two-dimensional plate deflections and that the bifurcation is
supercritical, the decreasing threshold $U^*_{d}$ should converge to the present theoretical predictions
(based on one-dimensional flutter modes). This is indeed what is observed in  figure~\ref{fig:criticalU}(a)
giving another argument for a supercritical bifurcation. 

Undesired two-dimensional plate deflections can also be driven by gravity for
very long or very light plates. This is what happens when the mass ratio is
too large causing the plate to sag, twist or bend under its own weight. In the present study, these undesired effects occur for
$M^* \gtrsim 5$ and this is why  the analysis has been restricted to $M^* <2$. 
When these effects are present however, the instability threshold cannot be predicted by
simple means and this 
presumably explains why the critical velocity measured in the experiments of
\cite{Watanabe2002a} and \cite{Yamaguchi2000} for very large $M^*$
is about an order of magnitude
larger than the predictions of a two-dimensional model.

In this paper, we have argued that the hysteresis loop could be caused by (undesired) 
two-dimensional plate deflections. We have also suggested that the bifurcation could be supercritical
contrary to most experimental works found in the literature.  
Our arguments are not solid proof however, and the nature of the bifurcation remains an open
question that still need further studies.

\begin{acknowledgments}
The authors acknowledge support from the French ANR (No. ANR-06-JCJC-0087).
\end{acknowledgments}

\bibliographystyle{jfm}
\bibliography{/Users/Ch/Documents/LaTeX/biblio}

\end{document}